\documentclass[aip,apl,reprint]{revtex4-1}
\usepackage[utf8]{inputenc}
\usepackage{graphicx}
\usepackage{siunitx}				
\usepackage{textcomp}
\usepackage{amsmath}
\begin{document}
\title{Holographic optical bottle beams}
\author{Christina Alpmann}
\email{c.alpmann@uni-muenster.de}
\author{Michael Esseling}
\author{Patrick Rose}
\author{Cornelia Denz}
\affiliation{Institut für Angewandte Physik and Center for Nonlinear Science (CeNoS), Westfälische Wilhelms-Universität Münster, Corrensstraße~2/4, 48149~Münster, Germany}
\begin{abstract}
We present a convolution approach for the generation of optical bottle beams that combines established techniques of holographic optical trapping with hollow intensity distributions in order to manipulate absorbing particles. The versatility of our method is demonstrated by the simultaneous stable trapping of multiple particles at defined positions. Furthermore, the presented phase shaping technique allows for the dynamic manipulation of absorbing particles along arbitrary paths.
\end{abstract}
\maketitle
The all-optical manipulation of transparent objects, such as colloidal particles or living cells, has evolved into a well-explored field with many applications in biology, chemistry, and medicine~\cite{Fazal2011}, whereas the trapping of absorbing particles with light has remained a challenge for a long time.

While the Paul trap uses electric fields to provide a confinement of charged and polar particles with particular importance in the field of atom optics and plasma physics, neutral absorbing particles might be trapped with light through photophoresis, originally described by Ehrenhaft~\cite{Ehrenhaft1917} in 1917. This force can be several magnitudes larger than scattering and gradient forces in conventional optical tweezers and results from thermal gradients on the particle's surface~\cite{Beresnev1993}. An inhomogeneous illumination of an absorbing particle leads to different surface temperatures and in turn to a spatially modulated vapor pressure of the surrounding medium. This pressure gradient is responsible for the photophoretic force which pushes absorbing particles away from regions of higher light intensity.

Stable three-dimensional trapping experiments with absorbing particles often used counter propagating vortex beams to counterbalance the additional scattering forces~\cite{Desyatnikov2009, Shvedov2009}. Recently, it has been shown that a stable manipulation is also possible with a single trapping beam of hollow intensity distribution~\cite{Shvedov2010}, often denoted as optical bottle beam~\cite{Arlt2000, Zhang2011, Shvedov2011}.

In order to increase the flexibility of this approach, it is highly desirable to generate multiple dynamic optical bottle beams at arbitrary positions. To achieve this aim, we introduce an approach based on a combination of known algorithms from the field of holographic optical tweezers with optical bottle beams.

As shown before, an optical bottle beam can be generated by superimposing two axially displaced optical vortices. In contrast to earlier approaches, which were based on interferometric techniques using either static holographic films~\cite{Arlt2000} or binary amplitude vortex holograms~\cite{Zhang2011}, we intend to apply pure but dynamic phase holograms. Therefore, it is convenient to write the complex field of such a bottle beam in a first step as the Fourier transform ($\mathcal{F}$) of its complex hologram as
\begin{equation}
E_{\mathrm{b}} = \mathcal{F} \left[ B \, \mathrm{e}^{\mathrm{i} \phi_{\mathrm{b}}} \right] \; ,
\label{eq:bottleField}
\end{equation}
where $B$ and $\phi_{\mathrm{b}}$ denote the (real) amplitude and phase distribution of the hologram, respectively.

For the superposition of two optical vortices with topological charge $m$, axially displaced by the bottle height $\Delta z$, this complex hologram in polar coordinates ($r$, $\varphi$) can be written as
\begin{equation}
B \, \mathrm{e}^{\mathrm{i} \phi_{\mathrm{b}}} = \mathrm{e}^{\mathrm{i} (m \varphi \, + \, f(\Delta z / 2) \, r^2)} + \mathrm{e}^{\mathrm{i} (m \varphi \, + \, f(-\Delta z / 2) \, r^2)}
\label{eq:bottleHologram}
\end{equation}
with $f(z) \, r^2$ being the phase of a Fresnel lens that leads to a focal shift to the axial position $z$. The phase $\phi_{\mathrm{b}}$ of this complex field finally gives the phase hologram of the desired bottle beam.

\begin{figure}
\centering
\includegraphics{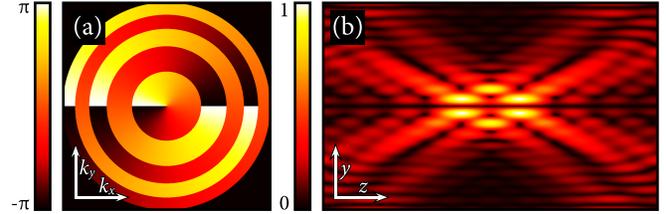}
\caption{(Color online) Optical bottle beam. (a)~Phase hologram of a bottle beam constituted of two axially displaced vortices with topological charge $m = 1$, (b) numerical simulation of corresponding bottle beam intensity distribution.}
\label{fig:bottleBeam}
\end{figure}

Figure~\ref{fig:bottleBeam}(a) depicts this phase hologram for the case $m = 1$, while Fig.~\ref{fig:bottleBeam}(b) shows a corresponding simulation of the intensity distribution after a Fourier transforming lens. This distribution clearly shows a dark central volume surrounded by regions of high light intensity necessary for the trapping of absorbing particles.

For practical applications, however, it is highly desirable to be able to generate multiple of those sophisticated traps and to control their transverse and axial positions individually on a dynamic basis. For conventional optical trapping, holographic optical tweezers provide a well-established solution to this demand. Initially realized with diffractive optical elements ~\cite{Dufresne1998} or nowadays with spatial light modulators~\cite{Liesener2000}, they allow for the generation of multiple dynamic foci in a given volume~\cite{Curtis2002}, by using combinations of Fresnel lenses and blazed gratings. The complex hologram output of the underlying algorithms, often denoted as ``prims and lenses'', can be written as
\begin{equation}
A \, \mathrm{e}^{\mathrm{i} \phi_{\mathrm{a}}} = \sum\limits_{n=1}^N \mathrm{e}^{\mathrm{i} (k_{x, n} \, r \cos \varphi \, + \, k_{y, n} \, r \sin \varphi \, + \, f(z_n) \, r^2)} \; ,
\label{eq:arrayHologram}
\end{equation}
where $N$ is the number of generated foci, $k_{x, n}$ and $k_{y, n}$ denote the transverse wave numbers of the respective blazed gratings, and $z_n$ give the axial focus positions. Again, a Fourier transform of this complex hologram yields the field generated in real space
\begin{equation}
E_{\mathrm{a}} = \mathcal{F} \left[ A \, \mathrm{e}^{\mathrm{i} \phi_{\mathrm{a}}} \right] \; .
\label{eq:arrayField}
\end{equation}

Since the multiple foci in this complex field represent a sum of three-dimensional $\delta$-distributions, a convolution ($\ast$) of~(\ref{eq:bottleField}) with~(\ref{eq:arrayField}) will give a field with $N$ bottle beams each located at a focus position of the prisms and lenses algorithm~\cite{Curtis2002}~\cite{Curtis2002}. Using the convolution theorem, this can be written as
\begin{align}
E_{\mathrm{ba}} &= E_{\mathrm{b}} \ast E_{\mathrm{a}} \nonumber \\
                &= \mathcal{F} \left[ B \, \mathrm{e}^{\mathrm{i} \phi_{\mathrm{b}}} \right] \ast \mathcal{F} \left[ A \, \mathrm{e}^{\mathrm{i} \phi_{\mathrm{a}}} \right] \nonumber \\
                &= \mathcal{F} \left[ B A \, \mathrm{e}^{\mathrm{i} (\phi_{\mathrm{b}} + \phi_{\mathrm{a}})} \right] \; ,
\label{eq:convolution}
\end{align}
and therefore the phase hologram of a field with multiple optical bottle beams is finally given by
\begin{equation}
\phi_{\mathrm{ba}} = (\phi_{\mathrm{b}} + \phi_{\mathrm{a}}) \!\!\!\! \mod 2\pi \; .
\label{eq:bottleArrayPhaseHologram}
\end{equation}

Since the holographic generation of point arrays has been studied in detail during the last years, many optimized algorithms exist that allow for a very fast calculation of the underlying phase holograms~\cite{Preece2009}. This facilitates even real-time tweezers applications for transparent particles~\cite{Hoerner2010}. According to Eq.~(\ref{eq:bottleArrayPhaseHologram}), such a sophisticated real-time tweezers setup can be modified to trap absorbing particles simply by adding the (constant) distribution $\phi_{\mathrm{b}}$ to each calculated phase hologram.

\begin{figure}
\centering
\includegraphics{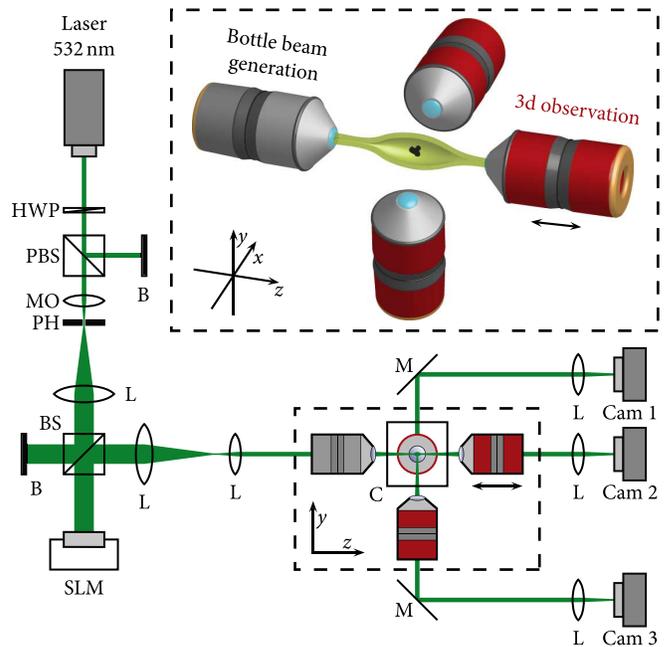}
\caption{(Color online) Schematic experimental setup for holographic optical bottle beam generation and three-dimensional particle observation. B: beam blocker, C: cuvette, HWP: half-wave plate, L: lens, M: mirror, MO: microscope objective, (P)BS: (polarizing) beam splitter, PH: pinhole, SLM: spatial light modulator.}
\label{fig:setup}
\end{figure}

To create the optical bottle beams for the three-dimensional manipulation of absorbing particles, we use the setup depicted in Fig.~\ref{fig:setup}. The beam of a frequency-doubled Nd:YAG laser is spatially filtered and expanded to illuminate a high-resolution phase-only spatial light modulator (SLM). The SLM, displaying a phase hologram calculated according to Eq.~(\ref{eq:bottleArrayPhaseHologram}) [cf.\ Fig.~\ref{fig:bottleBeam}(a)], is imaged to the back aperture of a 10x microscope objective which acts as a Fourier transforming lens to create the array of optical bottle beams. The focal volume of this objective is surrounded by a glass cuvette containing the particles to be manipulated. Once the hollow traps are generated, they are loaded by blowing particles into the stable trapping positions.

For the three-dimensional characterization of the bottle beams and for the observation of trapped particles, three microscopes are built around the trapping volume, covering the $xy$- (imaged by Cam~2), $xz$- (Cam~3), and $yz$-plane (Cam~1), respectively. In addition, the objective of the $xy$-microscope is mounted on a translation stage that allows for an axial scanning through the bottle beams. By stacking the $xy$-images obtained at different $z$-positions, axial $xz$- and $yz$-scans of bottle beams can be obtained.
 
In order to analyze the trapping, Cam~1 and Cam~3 simultaneously image the light scattered from trapped particles in the $xz$- and $yz$-plane, respectively. This allows for a precise three-dimensional determination of the particle positions.

We used home-made graphite flakes with a size distribution ranging from 2--\SI{20}{\micro\meter} as test particles. These particles fulfill the requirement of being large enough so that they cannot escape the optical bottle beam through its bottlenecks [cf.\ Fig.~\ref{fig:bottleBeam}(b)]. However, the photophoretic trapping should work with other absorbing particles as well, and we could transfer our method to, for instance, glassy carbon powder.

\begin{figure}
\centering
\includegraphics{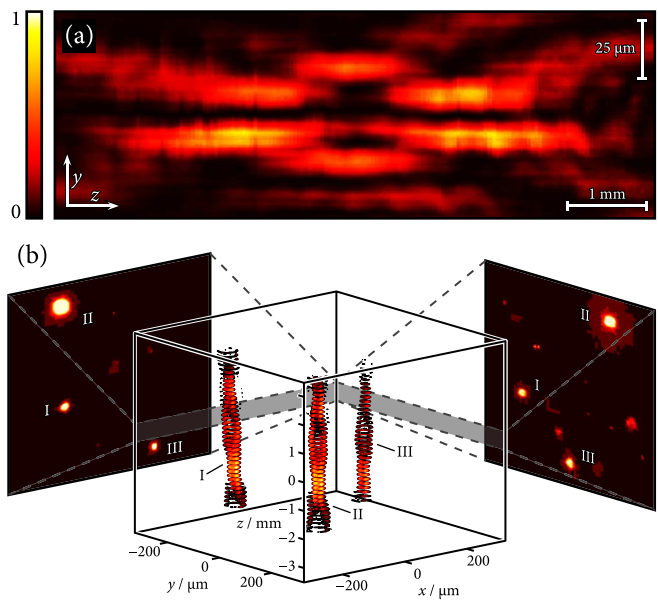}
\caption{(Color online) Experimental realization of optical bottle beams. (a)~$yz$-scan of the normalized intensity distribution of one bottle beam, (b)~simultaneous trapping of multiple absorbing particles. Stack of contour plots visualizes the three-dimensional intensity distribution of three bottle beams generated at different positions (center, enhanced online), side views show the light scattered by particles trapped in this configuration.}
\label{fig:3dConfig}
\end{figure}

Figure~\ref{fig:3dConfig}(a) shows a $yz$-scan of an optical bottle beam generated in our setup. This intensity distribution possesses regions of low intensity surrounded by high intensity areas as predicted by the numerical simulation depicted in Fig.~\ref{fig:bottleBeam}(b). The different length scales in $y$- and $z$-direction underline the asymmetric bottle shape which should lead to stable trapping in the transverse plane while trapped particles are expected to be less confined in $z$-direction.

The convolution approach [cf.\ Eq.~(\ref{eq:convolution})] enables us to simultaneously create multiple bottle beams at arbitrary transverse and longitudinal positions. Figure~\ref{fig:3dConfig}(b) illustrates a triangular configuration consisting of three bottle beams having their center at the ($x$, $y$, $z$) positions I~(\SI{-150}{\micro\meter}, \SI{-180}{\micro\meter}, \SI{100}{\micro\meter}), II~(\SI{-150}{\micro\meter}, \SI{160}{\micro\meter}, \SI{425}{\micro\meter}), and III~(\SI{150}{\micro\meter}, \SI{0}{\micro\meter}, \SI{-75}{\micro\meter}). The stack of contour plots verifies the complex three-dimensional trapping potential with all addressed bottle beams.

Absorbing graphite flakes are stably trapped in the desired positions using a laser power of about \SI{15}{\milli\watt} per bottle beam. The two side-view microscopic images in Fig.~\ref{fig:3dConfig}(b) show the particle positions in the $xz$- and $yz$-plane, respectively, which gives a precise three-dimensional verification of the addressed trapping potential. See supplementary material at [URL: http://
dx.doi.org/10.1063/1.3691957.1] for a three-dimensional animation of the measured intensity distribution.

\begin{figure}
\centering
\includegraphics{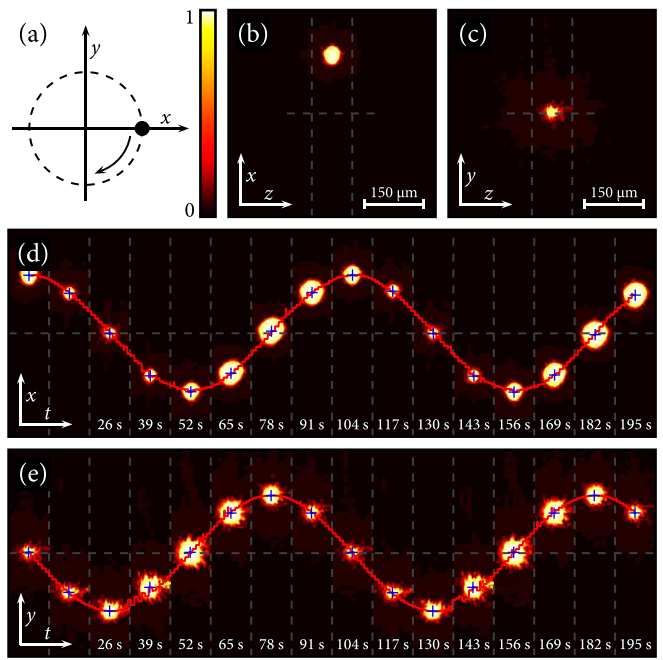}
\caption{(Color online) Dynamic movement of particles. (a)~Sketch of transverse circular motion, (b),~(c)~particle trapped at the start position observed in the $xz$- and $yz$-plane, respectively; dashed lines indicate particle centered cutouts for time series of particle movement depicted in~(d) and~(e), where crosses and lines show measured time-dependent particle positions in transverse direction (enhanced online).}
\label{fig:circularMovement}
\end{figure}

Besides this static positioning of multiple trapped particles in space, the dynamic change of particle configurations is possible with our holographic generation approach as well. To demonstrate this capability, we move a loaded trap on a circular path with a radius of \SI{150}{\micro\meter} around the optical axis. This movement is illustrated schematically in Fig.~\ref{fig:circularMovement}(a). Images of the particle, trapped in the bottle beam at the starting point, are shown in Figs.~\ref{fig:circularMovement}(b) and~\ref{fig:circularMovement}(c).

Respective images were taken during the whole circular movement of the particle, and Figs.~\ref{fig:circularMovement}(d) and~\ref{fig:circularMovement}(e) depict these time series. Here, each image cutout is centered around the $z$-position of the observed particle in order to account for the asymmetric bottle potential leading to a lower confinement in the axial direction. While the crosses mark the detected particle positions in the given snapshots, the underlying line gives the results of the position detection during the whole experiment.

As expected, the transverse displacement follows precisely a cosinusoidal and sinusoidal oscillation in time, respectively, and verifies an extremely robust particle confinement in both transverse directions. Over and above, the periodic change of the observed spot size in Figs.~\ref{fig:circularMovement}(d) and~\ref{fig:circularMovement}(e) confirms the oscillatory movement of the particle in and out of the microscopes' focal planes. See supplementary material at [URL: http://dx.doi.org/10.1063/1.3691957.2] for $xz$- and $yz$-videos of the moving particle.

In conclusion, we have demonstrated a convolution approach that enables the independent control of multiple complex trapping potentials. This approach is capable of creating dynamic bottle beam arrays to manipulate absorbing particles in a macroscopic trapping volume only by phase-shaping techniques and the number of simultaneously generated traps is only limited by the available laser power. Due to the marginal calculation effort of the method, it should be easy to modify existing holographic tweezers setups in order to control absorbing particles as well. Furthermore, it is self-evident that the presented convolution approach is not limited to bottle beams but can be used to generate an array of arbitrary holographic field distributions.

\vspace{3mm}
This work was partially supported by the DFG Grant No.~TRR61.

\end{document}